\documentclass[pre,aps,twocolumn,showpacs,preprintnumbers,superscriptaddress]{revtex4}

\usepackage{graphicx}% Include figure files
\usepackage{dcolumn}% Align table columns on decimal point
\usepackage{bm}% bold math
\usepackage{amssymb}
\usepackage{pifont}
\usepackage{amsmath}
\usepackage{times}
\usepackage[nooneline]{subfigure}

\begin{document}

\title{Existence of hysteresis in the Kuramoto model with
bimodal frequency distributions}

\author{Diego Paz\'o}
\affiliation{Instituto de F\'{i}sica de Cantabria (IFCA), CSIC-Universidad de Cantabria, E-39005 Santander, Spain }
\author{Ernest Montbri\'o}
\affiliation{Computational Neuroscience Group, Department of Information and Communication Technologies, Universitat Pompeu Fabra, 08003 Barcelona, Spain}
\affiliation{Center for Neural Science, New York University. 
New York, NY 10012, USA}

\date{\today}

\begin{abstract}
We investigate the transition to synchronization in the Kuramoto model with 
bimodal distributions of the natural frequencies.
Previous studies have concluded that the model exhibits a hysteretic phase transition
if the bimodal distribution is close to a unimodal one,
due to the shallowness the central dip.
Here we show that proximity to the unimodal-bimodal border does not 
necessarily imply hysteresis
when the width, but not the depth, of the central dip tends to zero.
We draw this conclusion from a detailed study of the Kuramoto model 
with a suitable family of bimodal distributions.
\end{abstract}
\pacs{05.45.Xt} %Synchronization; coupled oscillators
\maketitle 

\section{Introduction}
Understanding the dynamics of large populations of heterogeneous self-sustained 
oscillatory units is of great interest because they occur in a wide range 
of natural phenomena and technological applications \cite{PRK01}. 
Often a macroscopic system 
self-organizes into a synchronous state, in which a certain fraction of its 
units acquires a  common frequency. This occurs as a consequence of the mutual 
interactions among the oscillators and despite the differences in their rhythms 
\cite{Win80}. Examples of collective synchronization include pacemaker cells in 
the heart and nervous system \cite{GM88,Dye91},  synchronously
flashing fireflies \cite{Buc88}, collective oscillations of pancreatic 
beta cells \cite{SR91} and pedestrian induced oscillations in bridges \cite{SAM+05}. 

A fundamental contribution to the study of collective synchronization
was the model proposed by Kuramoto \cite{Kur84}. This model, and a large number
of extensions of it, has been extensively studied because it is analytically tractable but still captures the essential dynamics of collective 
synchronization phenomena (for reviews see~\cite{Str00,PRK01,MMZ04,ABP+05}). 
The original Kuramoto model consists of a population of $N$ oscillators 
interacting all to all. The state of an oscillator $i$ is described by its 
phase $\theta_i(t)$ that evolves in time according to 
\begin{equation}
\dot{\theta}_i = \omega_i-\frac{K}{N} \sum_{j=1}^{N} \sin (\theta_i-\theta_j).
\label{model0}
\end{equation}
The parameter $K$ determines the strength of the interaction between 
one oscillator and another. The oscillators are considered to have different 
natural frequencies $\omega_i$, that are taken from a
probability distribution $g(\omega)$. In his analysis Kuramoto adopted
the thermodynamic limit $N\to\infty$ and considered $g(\omega)$ to be 
symmetric. In this case, and without loss of generality, the 
distribution can always be centered at zero, {\em i.e.}~$g(\omega)=g(-\omega)$, 
by going into a rotating  framework $\theta_j\to\theta_j+\Omega t$. 

Kuramoto found useful to study the synchronization dynamics of system
(\ref{model0}) in terms of a complex order parameter $z=N^{-1}\sum_{j=1}^N
\exp(i~\theta_j)$. Note that $z$ is a mean field that indicates the 
onset of coherence due to synchronization in the population.
System (\ref{model0}) possesses an incoherent state with $z=0$ (that exists for all values of the
coupling strength $K$) in which the oscillators rotate independently as if 
they were uncoupled, $\theta_i(t)\sim\omega_i t$.  
Using a self-consistency argument, Kuramoto found that for a unimodal 
distribution $g(\omega)$, above the coupling's critical value 
\begin{equation}
K_c=\frac{2}{\pi g(0)}, 
\label{Kc}
\end{equation}
a new solution with asymptotics 
\begin{equation}
|z|\approx {\frac{4}{K_c^2}} \sqrt{\frac{K-K_c}{-\pi g''(0)}} 
\label{z_asympt}
\end{equation}
branches off the incoherent ($z=0$) solution.
This emerging solution is a \emph{partially synchronized} (PS) state,
in which a subset of the population $\cal{S}$ entrains to the central frequency
($\theta_{i\in \cal{S}}={\rm const.}$).

Equation~(\ref{z_asympt}) shows that the orientation of the 
PS bifurcating branch depends on whether the distribution is concave or convex
at its center. As a consequence of that, at $K=K_c$ the PS state is expected to
bifurcate supercritically for unimodal distributions ($g''(0)<0$) and 
subcritically for bimodal distributions ($g''(0)>0$).
However, Kuramoto's 
analysis did not permit to study the stability of the solutions 
and thus one cannot conclude whether bimodal distributions show
bistability close to the transition point (\ref{Kc}) 
(see discussion in p.~75 of \cite{Kur84}). 
In fact, Kuramoto discarded the possibility of bistability.
Instead he expected the incoherent state to become unstable
earlier, {\em i.e.}~at a certain critical 
value $K'_c<K_c$, via the formation of two symmetric clusters of synchronized
oscillators near the distribution's maxima
(later Crawford called this state \emph{standing wave} (SW) \cite{Cra94}).
As the coupling is increased further, he predicted that the interaction 
between the clusters would tend to synchronize them forming a single synchronized 
group, {\em i.e.}~a PS state.

\subsection{Sum of unimodal distributions with different
mean}\label{SecSOU}

After Kuramoto's seminal work \cite{Kur84}, several articles have 
further investigated the synchronization transition in model (\ref{model0}) 
with symmetric bimodal distributions
\cite{BNS92,Cra94,BPS98,Bon00,MPS06,MBS+09}. 
These studies assumed $g(\omega)$ to be the superposition 
of two identical even unimodal distributions $\tilde g(\omega)$
centered at $\pm \omega_0$: $g(\omega)=\tilde g(\omega+\omega_0)+
\tilde g(\omega-\omega_0)$ \footnote{The choice $\tilde g$ to be of 
Lorentzian (Cauchy) type is popular because the mathematics usually simplifies.
Some works however investigate a population consisting of two
groups of identical oscillators [$\tilde g(\omega)=\delta(\omega)$] with model
(\ref{model0}) in the presence of noise \cite{BNS92,BPS98}.}.
Parameter $\omega_0$ controls the separation of the peaks.
Decreasing $\omega_0$ the distribution's maxima 
approach each other and, at the same time, the central distribution's 
dip becomes shallower ({\em i.e.} $g(0)$ increases). 
Eventually, 
at a value $\omega_0=\omega_{0B}$ that satisfies 
\begin{equation}
g''(\omega=0)|_{\omega_0=\omega_{0B}}=0.
\label{trans0}
\end{equation}
the peaks merge and the distribution becomes unimodal.
The dynamics of the Kuramoto model for distributions of this type is as follows \cite{MBS+09}:
When  the peaks are well separated ($\omega_0$ larger than a certain value $\omega_{0D}$)
the transitions increasing $K$ are as Kuramoto foresaw: Incoherence $\to$ SW $\to$ PS.
However, if the peaks are near ($\omega_{0D}>\omega_0 >\omega_{0B}$)
there exists a range of $K$ below $K_c$ where bistability between incoherence 
and either a PS or a SW state is observed, as Eq.~(\ref{z_asympt}) suggested 
\footnote{
Similar results have been obtained studying the interaction between populations
with Lorentzian frequency distributions \cite{OK91,MKB04,BHO+08}. In this context 
the bimodal distribution  arises naturally as the superposition the two 
unimodal distributions.}.

\subsection{Difference of unimodal distributions with different
width}\label{SecDOU}

%%%%%%%%%%%%%%%%%%%%%%%%%%%%%%%%%%%%%%%%%%%%%%%%%%%%%%%%%%%%%%%%%%%%%%%%%
\begin{figure}
\centerline{\includegraphics *[width=80mm,clip=true]{fig1.eps}}
\caption{(Color online) Examples of bimodal frequency distributions 
given by Eq.~(\ref{general}) with $\delta=1$. 
Left panel: $\xi=\gamma$ (what implies $g(0)=0$). Note that 
as $\gamma$ decreases the maxima of the distribution become closer.
For \emph{all} these distributions (with $\xi=\gamma$) the route to 
synchronization as $K$ is increased 
from zero  is I$\to$SW$\to$PS, c.f.~Fig.~\ref{phasediagram0}. 
Right panel: Two examples with $\xi < \gamma$. 
The distribution depicted with a continuous line has well
separated peaks and shows a transition  I$\to$SW$\to$PS, whereas the other 
distribution is closer to the unimodal limit (\ref{xiB}) and presents 
hysteresis in the route to synchronization, c.f.~Fig.~\ref{phasediagram1}.
\label{distributions}}
\end{figure}
%%%%%%%%%%%%%%%%%%%%%%%%%%%%%%%%%%%%%%%%%%%%%%%%%%%%%%%%%%%%%%%%%%%%%%%%

In this article we are interested in understanding the 
synchronization transition in the Kuramoto model with bimodal 
distributions in situations that cannot be achieved
summing even unimodal distributions. In particular summing even distributions
implies that if the peaks are brought closer the central dip becomes less deep
(unless the distributions are Dirac deltas).
Thus we cannot approach the peaks arbitrarily near while keeping
the central dip's depth (see {\em e.g.} in the left panel of Fig.~\ref{distributions}
for a distribution family with constant depth but arbitrary
distance between the peaks).

We will use a
family of bimodal distributions that are constructed as
the difference of two unimodal even
functions with the same mean and different widths:
$g(\omega)=\tilde{g}_1(\omega)-\tilde{g}_2(\omega)$.
These distributions could be useful to model systems in which a fraction of the central
natural frequencies of a population $\tilde{g}_1$ is missing due to  
for example, some  
resonance, symmetry, or external disturbance.

We choose the functions $\tilde{g}_i$ to be Lorentzians, because 
of their mathematical tractability. 
Assuming $\delta >\gamma$ the normalized distribution reads
\begin{equation}
g(\omega)=\frac{\Xi}{\pi}
\left[\frac{\delta^2}{\omega^2+\delta^2}-\xi\left(
\frac{\gamma}{\omega^2+\gamma^2}\right)\right]
\label{general}
\end{equation}
with $\xi \le \gamma $ to be well defined, and $\Xi=1/(\delta-\xi)$ is the 
normalization constant. Without loss of generality we assume $\delta=1$ hereafter,
because this can be always achieved rescaling $\omega$, time
and the parameters: 
$\omega'=\omega/\delta$, $t'=t \delta$,
$K'=K/\delta$, $\gamma'=\gamma/\delta$ and $\xi'=\xi/\delta$. 
We will also drop the primes to lighten the notation.
Figure~\ref{distributions} shows several examples of
distributions (\ref{general}).
Distribution family (\ref{general})
can exhibit an arbitrarily deep minimum while keeping 
the maxima as near as wished.

\begin{figure}
\centerline{\includegraphics*[width=80mm,clip=true]{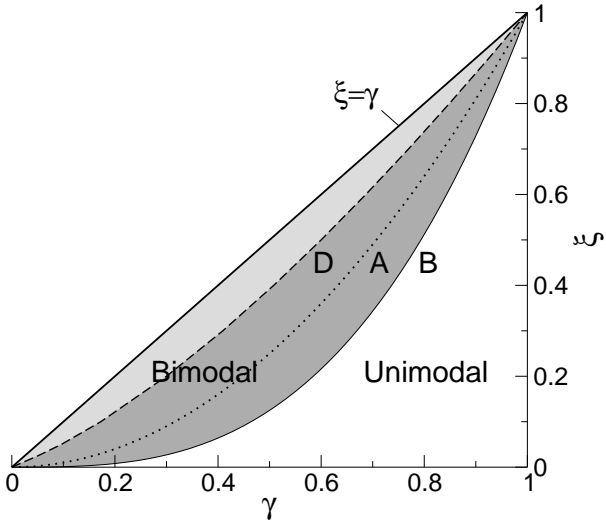}}
\caption{The parameter space of distribution (\ref{general}) 
[not defined above the bisectrix $\xi=\gamma$ neither at 
point (1,1)]. Function (\ref{general}) is unimodal below 
line B and bimodal above it (shaded regions). Three lines signal the loci 
of codimension-two bifurcations (A, B, and D) projected on the 
$(\gamma,\xi)$ plane. Between lines D and B 
(dark grey region) the transition to synchronization 
involves hysteresis.  
\label{codim2}}
\end{figure}

The left panel of Fig.~\ref{distributions} shows two examples 
for the case $\xi=\gamma$, which will be analyzed in detail below.
This case implies $g(0)=0$, which corresponds to the maximal value
of the ratio $\xi/\gamma=1$.
As $\gamma \to 0$, 
the central dip becomes infinitely narrow and at $\gamma=0$ 
the distribution becomes unimodal. 
This unimodal transition is therefore discontinuous and
satisfies \footnote{$g''(0)\sim \xi/\gamma^3$ diverges as $\xi\le\gamma \to 0$
if $\xi=O(\gamma^a)$ with $a<3$, {\em e.g.}~$\xi=\gamma$ ($a=1$).}:
\begin{equation}
\lim_{\gamma \to 0^+} g''(\omega=0)=\infty,
\label{trans1}
\end{equation}
In addition, distribution (\ref{general}) also presents the regular 
unimodal-bimodal border via $g''(0)=0$ at
%s
\begin{equation}
\xi_B= \gamma^3 
\label{xiB}
\end{equation}
with $\gamma\ne 0$ (line B in Fig.~\ref{codim2}).\\

The outline of the paper is as follows: 
Section~\ref{SecTh} summarizes recent theoretical results that permit to 
reduce the Kuramoto model to a system of ordinary differential equations 
with complex variables. These results are then used to find the two ODEs 
that describe the dynamics of the Kuramoto model with distribution
(\ref{general}).
In Sec.~\ref{sec_0} we study the special case $\xi=\gamma$, 
and we show that there indeed exists a transition to synchronization in 
absence of hysteresis independent on the separation between the 
distribution's maxima. Namely, in this case the route to synchronization 
is always: Incoherence$\to$SW$\to$PS. In Sec.~\ref{sec_1} 
we study the most general case $g(0)>0$, and determine the disposition
of the different synchronization scenarios with respect to the unimodal-bimodal border.

%%%%%%%%%%%%%%%%%%%%%%%%%%%%%%%%%%%%%%%%%%%%%%%%%%%%%%%%%%%%%%%%%%%%%%%%%
%%%%%%%%%%%%%%%%%%%%%%%%%%%%%%%%%%%%%%%%%%%%%%%%%%%%%%%%%%%%%%%%%%%%%%%%

\section{Low dimensional description of the Kuramoto model}\label{SecTh}

We start considering the thermodynamic limit
$N\rightarrow \infty$ of model 
(\ref{model0}). We drop hence the indices in
Eq.~(\ref{model0}) and introduce the probability density for the phases
$f(\theta,\omega,t)$~\cite{Kur84,SM91}. 
Then $f(\theta,\omega,t)\, d\theta \, d\omega$
represents the ratio of oscillators with phases between 
$\theta$ and $\theta+d\theta$,
and natural frequencies between $\omega$ and $\omega+d\omega$.
The density function $f$ obeys the continuity equation 
\begin{equation}
 \frac{\partial f}{\partial t}= - \frac{\partial(f v)}{\partial \theta},
\label{cont_eq}
\end{equation}
where, the angular velocity of the oscillators $v$ is given by
\begin{equation}
v(\theta, \omega, t)=\omega - K \int_0^{2 \pi} f(\theta', \omega, t) \sin(\theta-\theta ') d\theta'
\end{equation}
In the continuous formalism, the complex order parameter defined by 
Kuramoto becomes
\begin{equation}
z(t)=\int_{-\infty}^{\infty} \int_0^{2\pi} e^{i\theta} f(\theta,\omega,t) \, d\theta \, d\omega.
\label{z}
\end{equation}
Since the density function $f(\theta,\omega,t)$ is real and $2\pi$
periodic in the  $\theta$ variable, it admits the Fourier expansion
\begin{equation}
f(\theta,\omega,t)=\frac{g(\omega)}{2\pi}\left[ 1
+\sum_{n=1}^\infty \left( f_n(\omega,t) e^{in\theta} + {\rm c.c.} \right)\right],
\label{fourier}
\end{equation}
where $f_n=f_{-n}^*$. Note that the order parameter (\ref{z}) now reduces to
\begin{equation}
z^*(t)=\int_{-\infty}^{\infty} g(\omega) f_{1}(\omega,t) \, d\omega.
\label{z1}
\end{equation}
Substituting the Fourier series (\ref{fourier}) into the continuity equation
(\ref{cont_eq}), and using Eq.~(\ref{z1}) one gets an infinite set of
integro-differential equations for the Fourier modes
\begin{equation}
\dot{f}_n=-i n \omega  f_n + \frac{n K}{2} \left( z^* f_{n-1}-z  f_{n+1} \right) .
\label{fourier_set}
\end{equation}
Recently Ott and Antonsen (OA) found a very remarkable result~\cite{OA08}: The
ansatz
\begin{equation}
f_n(\omega,t)=\alpha(\omega,t)^n
\label{ansatz}
\end{equation}
is a particular --and usually the asymptotic-- solution of the
infinite set of Eqs.~(\ref{fourier_set})
if $\alpha$ satisfies
\begin{equation}
\dot \alpha=-i  \omega  \alpha+ \frac{K}{2} \left( z^*-z \alpha^2 \right) .
\label{alpha}
\end{equation}
Equation (\ref{alpha}) reduces to a finite set of ODEs
for distributions $g(\omega)$ with a finite set of simple poles
out of the real axis. Recalling $f_1=\alpha$ the order parameter
can be calculated by extending the integral in (\ref{z1}) to a contour
integration in the complex plane. This is possible
since $\alpha$ has an analytic continuation
in the lower half $\omega$-plane~\cite{OA08}.
In turn only the values of $\alpha$ at the poles of $g(\omega)$ with
negative imaginary part are relevant.

Several recent studies show that the ansatz (\ref{ansatz})
yields predictions in agreement with numerical 
simulations \cite{OA08,MBS+09,CS08,AFG+08,AMS+08,Lai09,LOA09}.
In addition Ott and Antonsen theoretically support the validity of their ansatz
for the case of a Lorentzian distribution \cite{OA09}.
So far, disagreement between the OA ansatz and numerical results
has been shown for frequency distributions with no spread and non-odd-symmetric
coupling function. This 
entails the freedom to select arbitrary values for some constants of
motion \cite{PR08}.

\subsection{Main Equations}

In this section we use the OA ansatz considering frequency
distribution (\ref{general}). This yields
two ODEs governing the dynamics inside the low-dimensional OA manifold.
First of all, it is convenient to express (\ref{general}) in partial fractions:
\begin{equation}
g(\omega)=\frac{\Xi}{2 \pi i} \left( \frac{1}{\omega-i}
-\frac{1}{\omega+i}  \right. 
 - \left.\frac{\xi}{\omega-\gamma i}
+\frac{\xi}{\omega+\gamma i} 
\right) .
\end{equation}
Then, according to Eq.~(\ref{z1}) the order parameter reads
\begin{equation}
z^*(t) = \Xi[ \alpha_1(t) - \xi\alpha_2(t)],
\label{zg}
\end{equation}
with $\alpha_{1}(t)=\alpha(\omega=-i,t)$, and $\alpha_{2}(t)=\alpha(\omega=-i\gamma,t)$.
Using (\ref{zg}) in Eq.~(\ref{alpha}), we obtain the following
two ODEs with complex variables
that govern the evolution of the order parameter (\ref{zg})
\begin{subequations}
\label{alphae}
\begin{equation}
\dot\alpha_1 =-\alpha_1 +k(\alpha_1-\xi \alpha_2) - k(\alpha_1^* - \xi \alpha_2^*) \alpha_1^2 \label{alpha1}
\end{equation}
\begin{equation}
\dot\alpha_2 =-\gamma\alpha_2 + k (\alpha_1-\xi\alpha_2) - k(\alpha_1^* - \xi \alpha_2^*) \alpha_2^2,
\label{alpha2}
\end{equation}
\end{subequations}
with $k=\Xi K/2$.
The phase space of Eqs.~(\ref{alphae}) is
four dimensional, but due to the global phase shift invariance
$(\alpha_1,\alpha_2)\to(\alpha_1 e^{i\beta},\alpha_2 e^{i\beta})$
the dynamics is actually three dimensional [see also Eqs.~(\ref{polar}) in Appendix A].

\subsection{Fixed points}
\label{fps}
According to Eq.~(\ref{zg}), the fixed points of Eqs.~(\ref{alphae}) correspond to
steady states of the order parameter $z$.
The trivial solution $\alpha_1=\alpha_2=0$ yields $z=0$, corresponding to
the incoherent state. 

In order to calculate the non-trivial fixed points, note first that invariance under
the action of the global rotation $e^{i\beta}$ allows us to
choose $\alpha_1=x_1+i y_1$ real, {\em i.e.} $\alpha_1=x_1$.
It follows from Eq.~(\ref{alpha1}) that the fixed points lie on the
subspace where $\alpha_2$ is real too.
We can therefore take $\alpha_1$ and $\alpha_2$ as real (keeping in mind
that a continuous of fixed points is generated under the action the neutral rotation $e^{i\beta}$). Hence, the equations for the fixed points are:
\begin{subequations}
\begin{equation}
0=-x_1+k(x_1- \xi x_2)(1-x_1^2)
\label{despxb0}
\end{equation}
\begin{equation}
0=-\gamma x_2+k (x_1-\xi x_2)(1-x_2^2)
\label{despyb0}
\end{equation}
\end{subequations}
Additionally, note that these equations are symmetric under the 
reflection $(x_1,x_2)\to (-x_1,-x_2)$. This implies that 
the solutions (with the exception of the solution at the origin) exist 
always in pairs with opposite signs $(\pm x_1, \pm x_2)$.

Subtracting Eq.~(\ref{despxb0}) from
Eq.~(\ref{despyb0}) multiplied by $\tfrac{\xi}{\gamma}$, we 
obtain 
$x_2^2=\tfrac{\gamma}{\xi}[ x_1^2+\tfrac{1}{k}+\tfrac{\xi}{\gamma}-1]$.
This can be substituted back into Eq.~(\ref{despxb0})
to get a cubic equation in $X\equiv x_1^2$:
\begin{eqnarray}
P(X)&=& k^2(1-\gamma\xi)X^3 \nonumber\\
&-&k\left[(2k-1)(1-\gamma\xi)-1+k\xi(\xi-\gamma)\right]X^2 \nonumber\\
&+&\left[(k^2-2k)(1-\gamma\xi)+1+2 k^2\xi(\xi-\gamma)  \right]X \nonumber\\
&-&k\xi\left[\gamma+k(\xi-\gamma)\right]=0 
\label{cubic2}
\end{eqnarray}
Each of the solutions of this equation yields two twin solutions with coordinates
\begin{equation}
x_1=\pm\sqrt{X}  \qquad \xi x_2=x_1[1-\tfrac{1}{k(1-X)}] .
\label{x1x2}
\end{equation}
After some algebra we obtain the relation of the solutions with order parameter:
\begin{equation}
|z|= \frac{2 \, \xi \,\sqrt{X}}{K(1-X)}.
\label{zmod}
\end{equation}
A steady state $(x_1,x_2)$ results in a time-independent value of $z$ and 
hence it should correspond to a partially synchronized state. 
However, note that $X$ can only take values within the range $X\in[0,
1-2\tfrac{\xi}{K}(\sqrt{\tfrac{\xi^2}{K^2}+1}-\tfrac{\xi}{K})]$ to have a $z$ value
consistent with its definition, {\em i.e.}~ $|z|\in [0,1]$. 

As the polynomial in Eq.~(\ref{cubic2}) is cubic, there is one real
solution, $X_{(3)}$, for all the parameters values.
This solution lays in the range $[0,1]$ (for $k>1$ a better bound is
$[1-1/k,1]$, since $P(1-1/k)=-\xi^2 < 0$ and $P(1)=1>0$).
However, it turns out that 
the fixed points associated to  $X_{(3)}$ are `unphysical' (even though in some
parameter 
ranges $|z|<1$). 
The reason 
is that the $x_2$ coordinate, corresponding to the solution $X_{(3)}$,
is always larger than 1 in absolute value. This implies $|\alpha_2|>1$,
and according to Eq.~(\ref{ansatz}) the Fourier series of the density function 
$f(\theta,\omega,t)$ is divergent at $\omega=-i\gamma$. 

We will see below that for large enough values of $K$ there exist two more
real solutions of $P(X)$: $X_{(1)}\le X_{(2)}<1-1/k$. In this case 
(except when $X_{(1)}$ becomes negative) such solutions indeed correspond 
to PS states of the original Kuramoto model (\ref{model0}).

%%%%%%%%%%%%%%%%%%%%%%%%%%%%%%%%%%%%%%%%%%%%%%%%%%%%%%%%%%%%%%%%%%%%%%%%%%
\section{Bimodal distributions vanishing at their center ($\xi=\gamma)$}
\label{sec_0}

In this section we consider $\xi=\gamma$ what implies that distribution
(\ref{general})
vanishes at its center, $g(0)=0$. In this case $\gamma$ (or $\xi$) 
becomes the parameter controlling the width of the central dip
of $g(\omega)$, and the maxima of the 
distribution are located at (see Fig.~\ref{distributions}, left panel):
\begin{equation}
\omega=\pm \gamma.
\label{maxs}
\end{equation}

\subsection{Stability of the incoherent state}

\begin{figure}
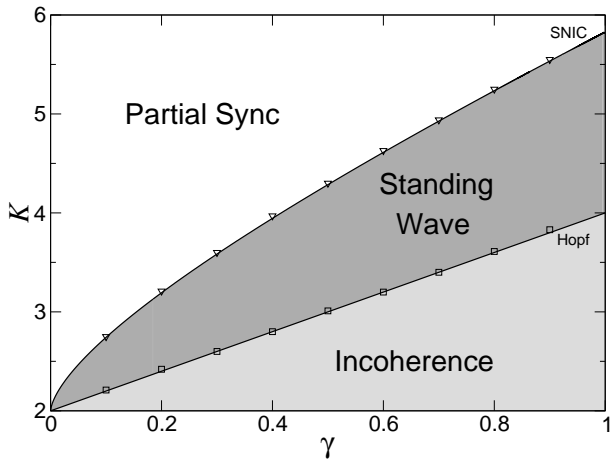

 \centerline{\includegraphics *[width=80mm]{fig3.eps}}
\caption{ Phase diagram for $\xi=\gamma$. For this case the synchronization
transition never involves hysteresis. The solid lines mark the 
saddle-node (SNIC) [from Eq.~(\ref{sn0})] and the Hopf [Eq.~(\ref{hopf})]
bifurcations. 
Symbols correspond to the numerical estimation of the bifurcation lines
via numerical integration of the original Eq.~(\ref{model0}) with $N=2000$.}
\label{phasediagram0}
\end{figure}

In the incoherent state the oscillators are uniformly distributed
in the interval $[0,2\pi)$, and thus the order parameter vanishes. 
This state corresponds to the
fixed point at the origin $\alpha_1=\alpha_2=0$.
%Now 
A linear 
stability analysis of Eqs.~(\ref{alphae}) reveals that this fixed point 
undergoes a degenerate Hopf bifurcation at $k_H=(1+\gamma)/(1-\gamma)$. 
In terms of the original coupling constant $K$, we find 
\begin{equation}
K_H=2+2\gamma.
\label{hopf}
\end{equation}
At this point the eigenvalues are imaginary 
$\lambda_{1,2}=\lambda_{3,4}^*=i\sqrt{\gamma}$ and two-fold degenerate.
Observe that as $\gamma\to0$, the critical coupling
for a (unimodal) Lorentzian distribution of unit width is recovered:
$K_H(\gamma\to 0)=K_c=2/(\pi g(0))=2$. Figure \ref{phasediagram0} 
shows the boundary $K_H$ in the $(\gamma,K)$ plane. As expected, 
we find that as the central dip of the distribution broadens
(increasing $\gamma$) the
stability region of the incoherent state grows.

\subsection{Saddle-node bifurcation}

The cubic equation~(\ref{cubic2}) for the non-trivial fixed points becomes
greately simplified under the assumption $\xi=\gamma$:
\begin{eqnarray}
Q(X)&=&k^2(1-\gamma^2)X^3-k\left[(2k-1)(1-\gamma^2)-1\right]X^2 \nonumber\\
&+&\left[(k^2-2k)(1-\gamma^2)+1\right]X-\gamma^2 k=0 .
\label{cubic}
\end{eqnarray}
For $\gamma=0$ the central dip vanishes, and we recover the solutions 
for a Lorentzian distribution $X=0,1-1/k$.
When $\gamma>0$ 
there is a saddle-node bifurcation at
$k=k_{SN}$, {\em i.e.}
there is a transition from one (for $k<k_{SN}$)
to three solutions (for $k>k_{SN}$). 
%In the language of bifurcation theory this is a 
 $k_{SN}$ and $\gamma$ can be related
imposing the condition that the discriminant of $Q(X)$ vanishes. This gives
the following relation:
\begin{equation}
\gamma^2=\frac{8 k_{SN}^4 - (1 + 8 k_{SN}^2)^{3/2} + 20 k_{SN}^2 -1}
{8 k_{SN} (k_{SN} +1 )^3}.
\label{sn0}
\end{equation}
There are two important asymptotic values for this bifurcation line,
which expressed in terms of the original coupling constant $K$ are
\begin{eqnarray}
 K_{SN}(\gamma\to
0)=2+6\left(\frac{\gamma}{2}\right)^{2/3}+O(\gamma), \label{asymp0}\\
K_{SN}(\gamma\to 1)\simeq (3+\sqrt{8})\left( 1 - \tfrac{1-\gamma}{2} \right).
\label{asymp1}
\end{eqnarray}
When $K$ increases above $K_{SN}$ the born solutions
depart from each other $X_{(2)}-X_{(1)} \sim \sqrt{K-K_{SN}}+ \mathrm{h.o.t.}$
One solution becomes progressively
smaller ($dX_{(1)}(K)/dK <0$), whereas the second one grows ($dX_{(2)}(K)/dK
>0$). The latter solution $X_{(2)}$ yields a monotonically growing value of
$|z|$ with $K$.
This is not surprising because in the Kuramoto model, 
at large values of $K$, there exists always a stable 
PS solution with $d|z|/dK>0$
(and $\lim_{K\to\infty} |z| = 1$, {\em i.e.}~full synchronization).
We advance that the corresponding twin fixed points from $X_{(2)}$ are stable,
whereas the fixed points corresponding to $X_{(1)}$ are saddle.

%%%%%%%%%%%%%%%%%%%%%%%%%%%%%%%%%%%%%%%%%%%%%%%%%%%%%%%%%%%%%%%%%%%%%%%%%
%%%%%%%%%%%%%%%%%%%%%%%%%%%%%%%%%%%%%%%%%%%%%%%%%%%%%%%%%%%%%%%%%%%%%%%%

\subsection{Numerical simulations and phase diagram}

In this section we construct the phase diagram with the loci of Hopf
and saddle-node bifurcations that we have obtained above. Numerical simulations
of the reduced Eqs.~(\ref{alphae}) were 
carried out and compared with the full model (\ref{model0}).
This permits to relate the dynamics of the variables $\alpha_{1,2}$ 
with the actual dynamical states of the Kuramoto model.

As already mentioned, the four-dimensional system (\ref{alphae}) is effectively
three-dimensional due to the existence of a neutral global rotation.
Interestingly the attractors of the model are apparently embedded into a {\em
two}-dimensional plane. Numerical simulations of Eqs.~(\ref{alphae}) using
arbitrary initial conditions show that the dynamics always collapses into a
plane which, by virtue of the neutral rotation $e^{i\beta}$, can be made
coincident with the $(x_1,x_2)$ plane, hereafter referred to as the ``real
plane". 
The stability against perturbations transversal to the real plane
(and not tangent to the global rotation) is difficult to prove analytically.
For the fixed point $X_{(2)}$ born at the saddle-node bifurcation, the stability against
transversal perturbations is proven in Appendix A.
Other attractors (limit cycle) are transversally stable according to our
numerical simulations.

Numerical simulations of the reduced Eqs.~(\ref{alphae}) with either 
real or complex variables, it is irrelevant, reveal that 
\begin{enumerate}
\item[(i)] The Hopf bifurcation at $K=K_H$ is supercritical and it gives rise to
a limit cycle around the origin. Due to the reflection symmetry of the equations
$z(t)$ vanishes twice per period [this occurs when
$\alpha_1=\gamma \alpha_2$, see Eq.~(\ref{zg})].
It is therefore reasonable to assume that 
the limit cycle corresponds to the SW state, for which the two counter-rotating
clusters of phase-locked oscillators are $\pi$ out of phase twice per period.
\item[(ii)] The oscillatory dynamics appearing at $K_H$ is destroyed at
$K=K_{SN}$ where twin saddle-node bifurcations give rise to twin pairs of fixed
points {\em on} the limit cycle. This bifurcation
is known as SNIC (saddle-node on the invariant circle), or SNIPER (saddle-node
infinite period). As $K$ approaches $K_{SN}$ from below the period of
$|z(t)|$ diverges due to the slowing down of the dynamics at
the twin bottlenecks anticipating the cease of oscillations via the 
(double) SNIC bifurcation. 
\end{enumerate}

Finally, numerical simulations of the full Kuramoto model~(\ref{model0}) confirm
the scenario I $\to$ SW $\to$ PS predicted by the reduced equations (\ref{alphae}).
We have numerically determined the boundaries of different behaviors:
Square symbols in 
Fig.~\ref{phasediagram0} are points in which 
the incoherent state loses stability leading to a SW state. 
Additionally, triangles indicate points where 
the order parameter becomes stationary.

\subsection{Concluding remarks}

Distribution (\ref{general}) with $\xi=\gamma$ becomes unimodal only for
$\gamma=0$. As
$\gamma\to 0$ the bimodal distribution tends to a unimodal,
but the limit is nonregular.
The remarkable point is that bistability is not
observed, even if the central dip is extremely narrow ($\gamma\to 0$) . This
is in sharp contrast with the scenario found when the
peaks are close to merge with $g''(0)\to 0^+$ at the usual 
unimodal-bimodal transition (see below).

Another interesting fact is that the counter-rotating clusters of the SW 
are born at the Hopf bifurcation (\ref{hopf}) with frequencies 
$\pm\sqrt{\gamma}$,
although the maxima of the distribution are located at $\pm \gamma$. 
This means that the relative shift between distribution's maxima and 
cluster frequencies at the onset of the SW diverges as 
$\gamma\to 0$. This is a consequence of the extreme asymmetry of the 
peaks in this limit.

\section{Bimodal distributions nonvanishing at their center ($\xi<\gamma$)}
\label{sec_1}

In this section we analyze the case $\xi<\gamma$, which is 
complementary to the one studied in the previous setion ($\xi=\gamma$).
Thus, in the present case we let $\xi$ and $\gamma$ to be independent 
of each other (see Fig.~\ref{codim2}). 
As we did in the previous section, we determine first the
local bifurcations of the fixed points, and then we summarize our findings 
in the ($\gamma,K$) phase plane together with the results obtained by numerical
integration of the reduced Eqs.~(\ref{alphae}) as well as of the full
Kuramoto model (\ref{model0}).

%%%%%%%%%%%%%%%%%%%%%%%%%%%%%%%%%%%%%%%%%%%%%%%%%%%%%%%%%%%%%%%%%%%%%%%%%
\begin{figure}
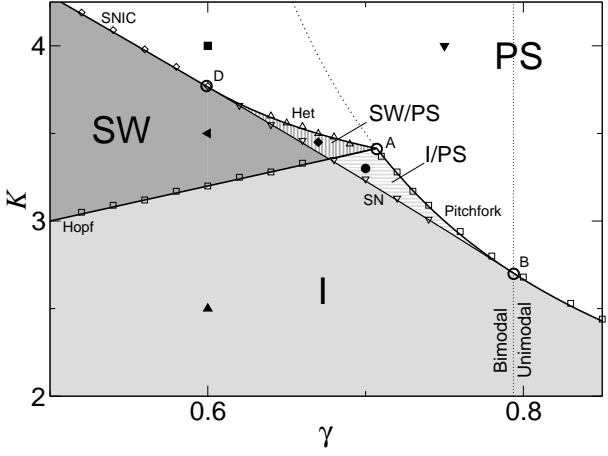

 \centerline{\includegraphics *[width=80mm]{fig4.eps}}
\caption{Phase diagram for $\xi=0.5$.
Solid lines mark the bifurcations: 
Saddle-node off the limit cycle (SN),
SNIC,  
Hopf bifurcation [Eq.~(\ref{hopf})], 
heteroclinic bifurcation (found numerically using the reduced equations),
and pitchfork bifurcation [Eq.~(\ref{pitchfork})]. 
Three big circles signal the codimension-two points:
(A) Takens-Bogdanov, (B) degenerate pitchfork,
(D) saddle-node separatrix-loop. The open symbols correspond to different 
bifurcations found by numerical integration of Eqs.~(\ref{model0})
with $N=2000$. Filled symbols inside each region indicate
parameter values for the phase portraits in Fig.~\ref{fig5}.}
\label{phasediagram1}
\end{figure}
%%%%%%%%%%%%%%%%%%%%%%%%%%%%%%%%%%%%%%%%%%%%%%%%%%%%%%%%%%%%%%%%%%%%%%%%

%%%%%%%%%%%%%%%%%%%%%%%%%%%%%%%%%%%%%%%%%%%%%%%%%%%%%%%%%%%%%%%%%%%%%%%%%
\begin{figure}
 \centerline{\includegraphics *[width=80mm]{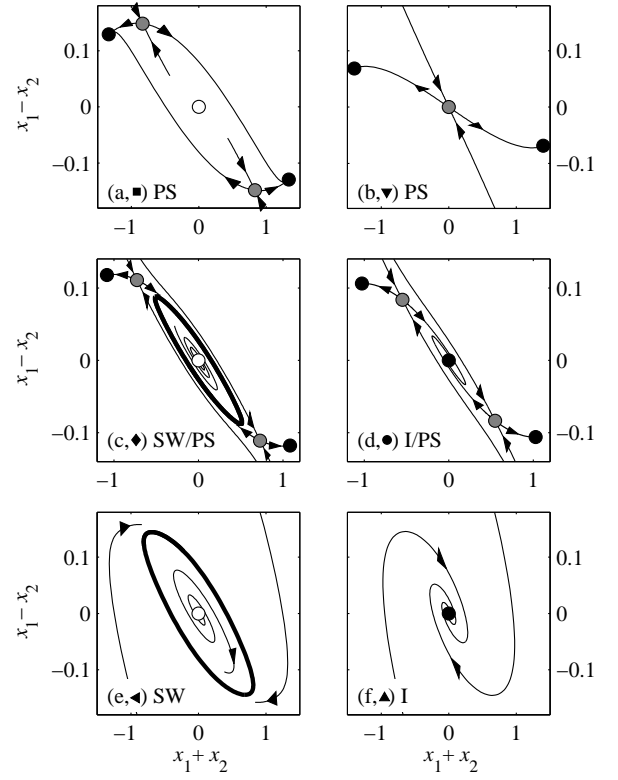}}
\caption{Phase portraits in (rotated) $x_1,x_2$ coordinates
for qualitatively different cases. Each panel corresponds
to a value of $\gamma$ and $K$ at the position of a filled symbol in Fig.~\ref{phasediagram1}.
(a,b) Partial synchronization with $K=4$, and (a) $\gamma=0.6$ and (b) $\gamma=0.75$;
(c) Coexistence SW/PS: $\gamma=0.67$, $K=3.45$;
(d) Coexistence I/PS: $\gamma=0.7$, $K=3.3$ ;
(e) SW, $\gamma=0.6$, $K=3.5$ ;
(f) I, $\gamma=0.6$, $K=2.5$.}
\label{fig5}
\end{figure}
%%%%%%%%%%%%%%%%%%%%%%%%%%%%%%%%%%%%%%%%%%%%%%%%%%%%%%%%%%%%%%%%%%%%%%%%

\subsection{Fixed points}

\subsubsection{The incoherent state and its stability}
The incoherent state becomes unstable in two possible ways depending on the
value of $\xi$ with respect to:
\begin{equation}
\xi_A=\gamma^2
\label{xia}
\end{equation}
(see line A in Fig.~\ref{codim2}).
For $\xi<\xi_A$, there is a degenerate Hopf bifurcation at
the critical value $K_H$ given by Eq.~(\ref{hopf}) which
%, surprisingly, 
is independent of $\xi$. For $\xi>\xi_A$, the instability
of the incoherent state occurs via a pitchfork bifurcation at:
\begin{equation}
K_P=\frac{2}{\pi g(0)}=\frac{2\gamma (1-\xi)}{\gamma-\xi}.
\label{pitchfork}
\end{equation}
The bifurcation is subcritical, and it switches to supercritical when the distribution
becomes unimodal at $\gamma > \xi_B^{1/3}$.
The loci of Hopf and pitchfork bifurcations collide at the codimension-two point
where $K_H=K_P$
and $\xi=\xi_A$.
This point is of the double zero eigenvalue type (Takens-Bogdanov) \cite{Guckenheimer}.

The boundaries (\ref{hopf}) and (\ref{pitchfork}) for Hopf and pitchfork
instabilities have been also obtained following a different approach in Appendix B.

\subsubsection{Non-trivial fixed points (partial synchronization)}

A saddle-node bifurcation occurs when $P(X)$ in Eq.~(\ref{cubic2}) has
exactly two roots (one of them two-fold degenerate). And this bifurcation point
can be determined numerically finding the value of $k$ where the discriminant of
$P(X)$ vanishes. The scenario is similar to the one observed for $\xi=\gamma$,
but in this case the saddle solution $X_{(1)}>0$ exists up to the pitchfork 
bifurcation with the origin at $K=K_P$. If the distribution is unimodal 
$X_{(1)}<0$ what makes this solution not valid.

\subsection{Numerical simulations and phase diagram}

Our analytical results provide information about local bifurcations. 
In addition we have performed numerical simulations of the 
ODEs (\ref{alphae}), in order to obtain the full 
system's picture. As occurred in the previous section, we can assume that 
$\alpha_j$ are real variables. In 
addition, we have performed numerical simulations of the original 
system that indicate that this assumption yields to correct results.

Figure~\ref{phasediagram1} shows the disposition of qualitatively different
dynamics in the parameters space spanned by $\gamma$ and $K$,
for a particular value of $\xi$.
Like in \cite{MBS+09} we find that
three codimension-two points organize
the parameter space:
Takens-Bogdanov (A), degenerate pitchfork (B), and 
Saddle-node separatrix-loop (D) \cite{schecter87}.
The three codimension-two points collapse at $\xi=\gamma=0$, see
Fig.~\ref{codim2}, and expressions (\ref{xia}) and (\ref{xiB}).
Line D approaches the origin linearly: 
$\xi_D (\gamma \to 0)= a \gamma$ with $a\simeq0.493$, 
suspiciously close to $\tfrac{1}{2}$.

One can better understand Fig.~\ref{phasediagram1}
looking at the panels of Fig.~\ref{fig5}, in which phase
portraits for qualitatively different states are shown.
In the rightmost part of Fig.~\ref{phasediagram1}, $\gamma > \gamma_B
=\xi^{1/3}$, the distribution becomes unimodal, and thus the standard 
route to partial synchronization is found. 
In the leftmost part, 
$\xi \le \gamma < \gamma_D\simeq 0.59997$ ($K_D\simeq3.7646$),
we have the same route than in the previous section, 
{\em i.e.}~a SW state limited by Hopf and SNIC bifurcations. 
In contrast, in the central part of the 
phase diagram (around point A), there exist 
two regions with bistability where the observed asymptotic state
depends on the initial conditions. In one region
(SW/PS) standing waves and
partial synchronization coexist, and the SW state (a limit cycle)
disappears via a heteroclinic collision with
the saddle points born at mirror saddle-node bifurcations.
In the second region (I/PS)
incoherence and partial synchronization coexist. 

Bifurcation lines in Fig.~\ref{phasediagram1} are calculated from analytical
results and from numerical integration of the ODEs (\ref{alphae}). Empty symbols
in the figure show the bifurcations determined integrating the Kuramoto model
with $N=2000$. The agreement is good and confirms the validity of the OA
ansatz.

\subsubsection{Codimension-two point A}

In this subsection we make a short digression about the codimension-two point
A and the importance of the symmetries in the model. 
Point A in Fig.~\ref{phasediagram1} is a Takens-Bogdanov point of system
(\ref{alphae}) that has $O(2)$ symmetry. This stems from the inherent $O(2)$
symmetry of the Kuramoto model [with symmetric $g(\omega)$].
Numerics show that the asymptotic dynamics occurs in the real
plane ---{\em i.e.}~Eqs.~(\ref{alphae}) with real coordinates--- where the symmetry
group is only $Z_2 \subset O(2)$.
This symmetry imposes the global bifurcation (Het) to be nontangent to the Hopf
line \cite{Guckenheimer},
in contrast with a nonsymmetric Takens-Bogdanov point.
Two scenarios are possible around the odd-symmetric Takens-Bogdanov
point \cite{Guckenheimer}. Hence, one may wonder if the
alternative 
scenario, involving a saddle-node bifurcation
of limit-cycles, might also be found in the Kuramoto model.

The  scenario that we have presented in this section (see also \cite{MBS+09}) is
apparently the same one Bonilla {\em et al.} \cite{BPS98} uncovered in the
neighborhood of the Takens-Bogdanov point for the Kuramoto model with additive
noise and a bi-delta frequency distribution. In that work the full $O(2)$
symmetry is taken into account.
Refs.~\cite{Cra94,BPS98} found that, due to the $O(2)$ symmetry, the degenerate 
Hopf bifurcation gives rise to a branch of unstable traveling wave solutions,
in addition to the stable SW. 
According to \cite{BPS98} these traveling wave solutions 
should disappear at a certain  $K<K_P$
in a local bifurcation with the saddle fixed points $X_{(1)}$
born at the SN bifurcations.
This bifurcation reverses the transversal stability
of the saddle fixed points,
what in turn makes congruent the pitchfork bifurcation of these fixed points
with the completely unstable fixed point at origin.
We think these traveling wave solutions and their associated
bifurcations are captured by the reduced Eqs.~(\ref{alphae}) because the OA ansatz has 
retained the $O(2)$ symmetry of the model. This means that although the relevant dynamics
(the attractors) are inside the real plane of $(\alpha_1,\alpha_2)$,
physical unstable objects (traveling waves) ``live" outside this plane.

\section{Conclusions}
\label{sec_final}

We have investigated the routes to synchronization in the Kuramoto model 
with a bimodal distribution constructed as the difference of two unimodal
distributions of different widths. 
These distributions admit an arbitrarily deep and narrow central dip,
what is not achievable in distribution types considered in the past.
This has allowed us to reinforce and extend the results recently published
in \cite{MBS+09}.

We have found that bimodal distributions (\ref{general}) near unimodality
produce hysteretic phase transitions,
except in some region in the neighborhood of
the unimodal limit $(\xi,\gamma)=(0,0)$, see Fig.~\ref{codim2}.

We expect a wide family of bimodal distributions
to exhibit the same qualitative features that Fig.~\ref{codim2}: 
The hysteretic region exist at the bimodal side of the unimodal-bimodal border,
and it shrinks as the nonregular unimodal-bimodal transition ($g''(0)=\infty$)
is approached.
Moreover the absence of hysteresis for $g(0)=0$
should be found in any bimodal distribution 
if the dependence is quadratic
---as in our distribution (\ref{general})--- or has a larger power:
$g(\omega)\propto|\omega|^\nu$ for small $\omega$,
with $\nu \ge 2$.

\acknowledgments
D.P.~acknowledges supports by CSIC under the Junta de Ampliaci\'on de
Estudios Programme (JAE-Doc), and by Ministerio de Educaci\'on
y Ciencia (Spain) under project No.~FIS2006-12253-C06-04. E.M.~acknowledges the financial support provided by the Centre de Recerca Matem\`atica (CRM), 08193
Bellaterra, Barcelona, Spain.

\appendix

\section{Proof of the transversal stability of fixed point $X_{(2)}$
in Sec.~\ref{sec_0} \label{appendixA}}

Global phase shift invariance,
$(\alpha_1,\alpha_2)\to(\alpha_1 e^{i\beta},\alpha_2 e^{i\beta})$,
allows to reduce Eqs.~(\ref{alphae}) 
in one dimension by passing to polar coordinates,
$\alpha_j=\rho_j e^{i\phi_j}$,
and defining the phase difference $\psi=\phi_1-\phi_2$.
We obtain three ODEs:
\begin{subequations}
\label{polar}
\begin{equation}
\dot\rho_1=-\rho_1+k(\rho_1-\xi\rho_2\cos\psi) (1-\rho_1^2)
\end{equation}
\begin{equation}
\dot\rho_2=-\gamma\rho_2 +k(\rho_1\cos\psi-\xi\rho_2)(1-\rho_2^2)
\end{equation}
\begin{equation}
\rho_1\rho_2\dot\psi = -k\left[(1-\xi)\rho_1^2\rho_2^2+\rho_1^2-
\xi\rho_2^2\right] \sin\psi \label{psi} 
\end{equation}
\end{subequations}

In Sec.~\ref{sec_0} we took $\xi=\gamma$ and found that twin saddle-node
bifurcations
(namely SNICs) give rise to two pairs of fixed points.
Here we prove (we rather sketch the proof) the transversal stability of the
mirror fixed points associated to $X_{(2)}$ via Eq.~(\ref{x1x2}).

First of all note that $X_{(2)}$ yields a fixed point ($x_1,x_2$), and its
mirror image, with $x_1$ and $x_2$ having the same sign, $\psi=0$. This is
a consequence of Eq.~(\ref{x1x2}) because $X_{(2)}<1-1/k$.
The latter inequality stems from the fact that $Q(1-1/k)=-\gamma^2 < 0$ and
by continuation of the solutions from $k=\infty$:
$\lim_{k\to\infty}X_{(1)}(k)=0$, $\lim_{k\to\infty}X_{(2,3)}(k)=1$.

Therefore we have to prove that factor
\begin{equation}
F=(1-\gamma)\rho_1^2\rho_2^2+\rho_1^2-\gamma\rho_2^2
\end{equation}
in Eq.~(\ref{psi}) for $\dot\psi$ is positive.
Replacing $\rho_1^2=X_{(2)}$ and $\rho_2^2=X_{(2)}+1/k$,
we have
\begin{equation}
F=(1-\gamma)[X_{(2)}^2+X_{(2)}(1+1/k)]-\gamma/k .
\end{equation}
As $X_{(2)}$ exists only above the saddle-node bifurcation ($k\ge k_{SN}$) and
$k_{SN}>k_H=(1+\gamma)/(1-\gamma)$.
\begin{equation}
F>(1-\gamma) h
\end{equation}
with
\begin{equation}
h =X_{(2)}^2+X_{(2)}-\gamma/(1+\gamma) .
\label{h}
\end{equation}
Then $h>0$ is a  sufficient condition for the transversal stability of the
fixed point.

It suffices to prove that $h$ is positive at the locus of the saddle-node
bifurcation because
$X_{(2)}(k,\gamma)$ exhibits its minimal value over $k$ precisely at the
bifurcation: $X_{(2)}(k>k_{SN},\gamma)> X_{(2)}(k_{SN},\gamma)$.
For our aim
it is better to parameterize the SNIC line by $k$ instead of $\gamma$. Hence we
to introduce in (\ref{h}) the expressions
\begin{enumerate}
\item[(i)]$\gamma$ as a function of $k_{SN}$, via Eq.~(\ref{sn0}).
\item[(ii)] $X_{(2)}(k_{SN})$, determined from (\ref{cubic}) in the two-fold root case.
\end{enumerate}
The calculation of terms (i) and (ii) can be readily
done with symbolic software  such as {\sc mathematica}.
As a result we obtain a function $h(k_{SN})$ that is positive in all the domain
of $k_{SN}\in(1,\infty)$.

Moreover using expressions (\ref{asymp0}) and (\ref{asymp1}) we can get
approximate expression for $h$ (as a function of $\gamma$):
\begin{eqnarray}
h(\gamma\to 0)= \left(\frac{\gamma}{2}\right)^{2/3} + O(\gamma)\\
h(\gamma\to 1) \simeq  0.0858
%0.0857864
\end{eqnarray}

\section{Stability of the incoherent state in the Kuramoto model with
noise\label{appendixB}}

For the sake of completeness, and as a double-check of some of the results
obtained, we study here the stability of the incoherent state
when the model is perturbed with additive white noises.
In this case, the right hand side of Eq.~(\ref{model0}) has to be
supplemented with uncorrelated fluctuating terms
$\eta_i$ satisfying $\left< \eta_i(t) \eta_j(t') \right> = 2\sigma \delta_{ij}
\delta(t-t')$.
So far a counterpart of the Ott-Antonsen ansatz for the stochastic problem has
not been found. It is nonetheless possible to obtain the stability boundary of
 incoherence resorting to the Strogatz and Mirollo relation for the 
discrete spectrum of eigenvalues $\lambda$ \cite{SM91}:
\begin{equation}
\frac{K}{2}\int_{-\infty}^{\infty} \frac{g(\omega)}{\lambda+\sigma+i\omega}
d\omega=1.
\label{incoherence}
\end{equation}
Considering the distribution of frequencies (\ref{general}), this equation
can be solved for the eigenvalues $\lambda$. 

Noise increases the domain of the incoherent state.
Hopf and pitchfork bifurcations continue
to occur, but the values of $K$ are shifted to larger values. We obtain:
\begin{eqnarray}
K_H &=& 2+2\gamma+4\sigma  \\
K_P &=& \frac{2(\gamma+\sigma)(1-\xi)(1+\sigma)}{(\gamma-\xi)+\sigma(1-\xi)}, 
\end{eqnarray}
that indeed reduce to Eqs.~(\ref{hopf}) and (\ref{pitchfork}) 
for $\sigma=0$.
The location of the Takens-Bogdanov point [c.f.~Eq.~(\ref{xia})]
also varies and now pitchfork and Hopf bifurcations collide
($K_H=K_P$) at :
\begin{equation}
\xi_{A}=\left( \frac{\gamma+\sigma}{1+\sigma } \right)^2 .
\end{equation}

\end{document}